# Normal epithelial and triple-negative breast cancer cells show the same invasion potential in rigid spatial confinement

Carlotta Ficorella[1,§], Rebeca Martínez Vázquez[2,§], Paul Heine[1], Eugenia Lepera[2], Jing Cao[3], Enrico Warmt[1], Roberto Osellame[2], Josef A. Käs[1]

[1] *Peter Debye Institute for Soft Matter Physics, University of Leipzig, Linnéstr. 5, 04103 Leipzig, Germany.*

[2] *Istituto di Fotonica e Nanotecnologie (IFN)-CNR3, P.zza Leonardo da Vinci 32, 20133 Milan, Italy.*

[3] *College of Chemistry and Molecular Engineering, Peking University, 100871, Beijing, P. R. China.*

§: Both authors contributed equally to this work.

Correspondence to Josef A. Käs: jkaes@physik.uni-leipzig.de; or Roberto Osellame: roberto.osellame@polimi.it

**ABSTRACT.**

The extra-cellular microenvironment has a fundamental role in tumor growth and progression, strongly affecting the migration strategies adopted by single cancer cells during metastatic invasion. In this study, we use a novel microfluidic device to investigate the ability of mesenchymal and epithelial breast tumor cells to fluidize and migrate through narrowing microstructures upon chemoattractant stimulation. We compare the migration behavior of two mesenchymal breast cancer cell lines and one epithelial cell line, and find that the epithelial cells are able to migrate through the narrowest microconstrictions as the more invasive mesenchymal cells. In addition, we demonstrate that migration of epithelial cells through a highly compressive environment can occur in absence of a chemoattractive stimulus, thus evidencing that they are just as prone to react to mechanical cues as invasive cells.

**Abbreviations used in this paper:** ECM – extra cellular matrix, 2PP – two photon polymerization, EMT – epithelial to mesenchymal transition, EGF – epithelial growth factor, GFP – green fluorescent protein, LOC – lab-on-a-chip, OS – optical stretcher.

## INTRODUCTION

Cells are the building blocks of an organism and, like solids, they have the ability to resist pulling and pushing forces in order to maintain tissue integrity. In skin, brain and other types of soft tissues, adherent cells and extracellular matrix (ECM) form a relatively elastic environment. During force generation and migration, motile cells, such as cancer cells, exploit very different migration strategies when invading the ECM, and can modify their mechanical properties to deform and flow when undergoing mechanical stress, transiting from a solid-like to a fluid-like behavior (Ahmed and Betz, 2015; Brábek et al., 2010; Lange and Fabry, 2013). For instance, during tumor invasion, cancer cells have the ability to acquire a migratory and invasive phenotype through the epithelial-to-mesenchymal transition (EMT), as well as to adapt their morphology to the surrounding ECM, generating protrusions to enter the endothelial cell barrier in order to intravasate into the lymphatic or circulatory systems and metastasize (Mierke, 2015). As different migration modes can be stimulated and regulated by extracellular cues, a factor that importantly affects tumor growth and progression is the extra-cellular micro-environment. It is observed that cancer cells respond to a variety of biochemical and mechanical signals in their micro-environment (Tse et al., 2012). Biochemical (or soluble) signals may result from factors such as the secretion of cytokines and growth factors (Bierie and Moses, 2006), and stresses due to a decrease of oxygen and glucose supply (Tse et al., 2012). Mechanical signals, on the other hand, may include ECM stiffening generated by collagen deposition and consistent thickening of collagen fibers (Acerbi et al., 2015), an increase in interstitial fluid pressure (Boucher and Jain, 1992), and compressive stresses due to confined tumor growth (Tse et al., 2012).

Mechanical compression plays an interesting role in the physics behind tumor development. It was already assumed that compression regulates the selection of metastatic cell populations or even stimulates tumor invasion, as it contributes to both genetic mutations and phenotypic changes that are related to malignancy (Tse et al., 2012). More recent studies have provided evidence that a combination of strong confinement and low adhesion may induce a transition from mesenchymal-like to amoeboid-like mode of migration (Liu et al., 2015). Previous research on cancer cell migration in confined geometries has mainly focused on the consequences of nuclear deformation and rupture. Deformation of the nucleus was observed during cell migration in collagen matrices and in subnuclear-sized microfluidic channels and pores of variable width and length (Mak et al., 2016; Paul et al., 2017; Pfeifer et al., 2018). Denais et al. (2016) have suggested that the nuclear envelope can rupture under strong confinement causing DNA damage and genomic instability, one of the hallmarks of cancer (Denais et al., 2016). An increase in intranuclear pressure, supposedly generated by the cytoskeleton, causes the nuclear membrane and envelope to rupture, hence, favoring an interchange of molecules between nucleoplasm and cytoplasm until the nuclear envelope is repaired (Isermann and Lammerding, 2017).

In the present study, we are investigating the dynamics of breast cancer cells in a hard-core confinement on a two-dimensional platform. Our main focus is on the internal engine which establishes the persistence and directionality of cell motion, the actin cytoskeleton, and its dynamical restructuring. In other words, we are interested in the ability of cells to alter their shape in order to achieve migration through narrowing spaces. Our microfluidic devices are created by laser-processing a SU-8 photoresist layer sandwiched between two glass substrates. Two photon polymerization (2PP) is a powerful technique for the fabrication of free-standing three dimensional (3D) polymer structures with micrometer-scale size (Maruo and Fourkas, 2008; Park et al., 2009). It is based on the irradiation of a photosensitive resin by a tightly focused femtosecond laser beam. Due to the high nonlinear nature of the absorption process, it enables the polymerization of the material in a small volume at the focal point (Baldacchini, 2018). Moving the sample relative to the laser beam it is possible to create arbitrary 3D polymerized structures with submicrometer resolution with almost no impact on the surrounding photoresist, which is then washed away by a development step. Mainly due to its high 3D resolution, 2PP is enabling several applications in fields like photonics, micro-mechanics, biology and lab-on-a-chip (LOC) (Baldacchini, 2018). In LOC applications, 2PP is particularly interesting because it allows the fabrication of 3D micrometer-sized structures in already closed microfluidic chips, by injecting the photosensitive resin in the microchannels and performing the laser processing through the top glass. In this way, complex structures can be embedded in microfluidic devices without the risk of destroying them during the closing process of the microfluidic channels (Amato et al., 2012; Wu et al., 2014). In this work, we are exploring this strategy to fabricate a chip for motility studies of cancer cells. The 3D capabilities of 2PP are exploited to fabricate microchannels with a tapered width going from a cavity larger than the cell dimensions to a much smaller confining one (~3µm). Such funnels provide the only connection between two reservoirs, fabricated in the glass substrate, thus directing the cells to move through the constrictions by following a chemoattractant stimulus. In the design of the device, we took care to achieve a clear optical access in the micro-funnel region in order to monitor the cell behavior during this crucial stage of their migration.

The rigidity of our narrowing microconstrictions offers the interesting opportunity to screen the deformability potential of cells, a fundamental ability in tumor invasion. The choice of a two-dimensional funnel geometry is motivated by the fact that it represents the simplest possible condition to observe how migrating cells react to compressive stress in narrowing confinement. Given the limited height of the constrictions (10 µm), and typical dimensions of cancer cells (~ 15 µm diameter, see Supplemetal Fig. 3) cells experienced confinement in all three spatial directions as they migrate into the funnel. Cells migrated in response to epithelial growth factor (EGF) chemoattraction and were driven into the narrowing confines of the funnels. The short-range hardcore repulsion experienced by single cells in the quasi-three-dimensional funnels restrained their motility, forcing them to adopt a polarized morphology in order to squeeze through the funnel opening. We demonstrate that strong confinement does not suppress motion in chemotactically stimulated MCF-10A epithelial cells, and that epithelial cells can be as responsive as invasive, mesenchymal MDA-MB-231 cells to stress induced by mechanical compression. Invasive cells switch to a bleb-based motility mode in order to squeeze through the highly constrictive aperture, whereas epithelial cells alternate between bleb and lamellipodial protrusions. We also observed that mesenchymal malignant MDA-MB-436 cells fail to switch their motility mode to optimize migration through the microconstrictions. Finally, we demonstrate that chemoattractant-based phenotypic changes in epithelial cells do not account for their ability to migrate in confining spaces.

## METHODS AND MATERIALS

**Cell culture**.

All human breast cancer cell lines used in this study were purchased by ATCC. MDA-MB-231 (Human breast adenocarcinoma, epithelial) and MDA-MB-436 cells (Human breast adenocarcinoma, derived from metastatic site) were cultured in DMEM (Biochrom, with 4500 mg/l glucose, L-glutamine, without sodium pyruvate) supplemented with 10% fetal calf serum FCS (Biochrom) and 1% 10000 U/ml penicillin/streptomycin (Biochrom). MCF-10A cells (Human breast, epithelial) were cultured in 1:1 mixture of DMEM and Ham's F12 medium (Biochrom, with L-Glutamine) supplemented with 5% Horse serum (HS), 20 ng/ml Epidermal Growth Factor (EGF), 10 μg/ml Insulin (Sigma-Aldrich, I6634 or I9278), 100 ng/ml Cholera Toxin (Sigma-Aldrich), 500 ng/ml Hydrocortisone (Sigma-Aldrich), and 1% 10000 U/ml penicillin/streptomycin (Biochrom). Cells were maintained at a temperature of 37°C and a 5% $CO_2$ concentration.

**Cell labeling.**

To label the F-actin, 1 μM staining solution was obtained by diluting 1:10000 Sir F-Actin stock solution (Spirochrome) with the appropriate medium. A pre-mixed solution with 0.1mM SiR-Actin and 0.1 mg/ml Hoechst-34580 (Invitrogen/Molecular Probes) to image the nuclei was also used, by diluting 1:1000 with appropriate medium. The premixed solution was obtained by diluting 5 μl Sir-Actin and 2 μl of 3.3 mg/ml Hoechst 34580 in 50 μl DMSO. Nuclear staining for Fig. 3(b) was obtained by diluting 1:10000 of Sir-DNA stock solution (Spirochrome) with appropriate medium.

**Chip fabrication and assembly.**

2PP allows for the fabrication of micrometer-sized channels in arbitrary geometries and with a high reproducibility. Consequently, it is relatively easy and fast to reconfigure the migration region depending on the desired test environment. We chose glass for the top and bottom layers due to its transparency; this gives us a device that enables imaging of cells, both in transmission and fluorescence microscopy. The chemical resistance of both glass and SU-8 make it easy to clean and reuse the device more than once. In the future we expect to exploit the 3D capabilities of 2PP to fabricate more complicated funnels geometries such as depth tapered funnels, thus forcing the squeezing of the cells in three dimensions. The fabrication pathway can be summarized in seven main steps (see Supplemental Fig. 1(b)):

**1 & 2. Fabrication of the chambers in the fused silica lid by femtosecond (fs) laser irradiation followed by chemical etching (FLICE) (Vishnubhatla et al., 2009).** Femtosecond laser irradiation is performed with the second harmonic (520 nm) of a femtosecond laser source (femtoREGEN, High Q) at a repetition rate of 1 MHz, with a pulse energy of 0.3 µJ. The laser beam is focused inside the fused silica substrate by a 20X, 0.45 numerical aperture (N.A.) microscope objective with 2 mm working distance (N-Achroplan, Zeiss). The irradiated pattern is made through a layer by layer irradiation, along the whole substrate depth for the passing holes and from 1 mm depth to the top of the glass for the reservoirs above (see Supplemental Fig. 1(a)). After the irradiation, the substrate is immersed for 4 hours in an ultrasonic bath of 20% HF aqueous solution at 35°C to selectively remove the irradiated volume. The passing holes delimit the area where the cells are able to migrate, while the upper reservoirs are made to increase the amount of liquid that can be used and consequently to reduce the effects of media evaporation during the motility experiments.

**3 & 4. Assembly of the three layers.** A 20 µm SU-8-3025 (Microchem) layer is spin-coated on a thin coverslip (Menzel-Gläser). Afterwards the fused silica lid is leaned on the SU-8 layer and the whole device undergoes a soft-baking process for 20 hours at 95°C. Due to this baking step the SU-8 film height is reduced to 10µm.

**5 & 6. In-chip 2PP and UV photolithography.** The fabrication of the micro-funnels is performed by 2PP of the SU-8 layer. Pulses from a fiber laser (Femtofiber pro NIR, Toptica), with 100 fs duration, 80 MHz repetition rate and 800 nm wavelength, are focused inside the SU-8 layer by a 63X, 0.75 N.A, microscope objective (LD-plan Neofluar, Zeiss) with the phase ring positioned to compensate for the thin glass presence. The sample is mounted onto a three-axis piezoelectric motion stage (Nanocube, P-611.3, Physik Instrumente), with nanometer resolution and 100 μm × 100 μm × 100 μm travel range, mounted on a long range xy linear stage (M-511.DD and M-605.2DD, Physik Instrumente). The 2PP is performed by irradiating the pattern plane by plane with a slicing step of 1.5 µm for the funnels and 3 µm for the boundary walls. The laser pulse energy and scanning speed are 0.375 nJ and 400 µm/s for the funnels and 0.5 nJ and 500 µm/s for the boundary walls. The specific layout of the funnels allows a precise control of the apertures and an optimal mechanical stability of the structures. The number of polymerized planes is such as to ensure adhesion of the SU8 structures to both the top and bottom glasses.

Immediately after the 2PP process a mask is positioned over the device covering the cell migration region, and the device is irradiated with an UV led (Hamamatsu). This allows the polymerization of a large frame around the area of interest, thus ensuring the perfect sealing of the device.

**7. Post-baking and development.** After the irradiation steps, a fast post-baking at 95°C for 5 minutes ensures a complete cross-linking of the polymer. Subsequently, the unirradiated photoresist is removed by immersing the chip in the SU-8 developer for 24 hours. A final immersion in a fresh solution of developer for some minutes ensures a complete removal of the unpolymerized photoresist in the narrower regions of the funnels.

**Experimental procedure.**

The optimal density (50% confluence) was assessed for each cell type with a EVE cell counter. For MDA-MB-231, MDA-MB-436 and MCF-10A cells a density of ca. $10^6$ cells/ml seeded on the culture chamber was chosen. A density value of approximately $5 \cdot 10^5$ cells/ml was enough for the MDA-MB-231 expressing GFP, since they tend to grow and divide much faster. In addition to seeding the cells directly on the glass surface of the chips, coating solutions such as poly-L-lysine (Sigma Aldrich, 0.01% [w/v], 50ml stored at 2-8°C), and fibronectin (stored in 1 ml aliquot, 50 µl/ml, at -80°C) were used in order to enhance cell motility and adhesion. 20 µl of coating solution was pipetted in both cell reservoirs. The chip was then incubated at 37°C for 1 hour (for fibronectin) or 2 hours (for poly-L-lysine), after which the coating solution was aspirated and the reservoirs rinsed with PBS. 20 µl of cell suspension was then added to the input chamber, and 20 µl of mixture of high-EGF medium was added in the output chamber. Further 30 µl of medium was added in each reservoir after 30 minutes.

During the measurements performed with a 10X or 20X objectives, the chips were placed in a 12-well plate with plastic bottom. The cells were observed for a maximum of four days, when the culture chamber population was too high to guarantee independent single cell squeezing through the constrictions. During these four days, the medium was refilled every 12 hours. For the experiments performed with higher magnification objectives the chip was placed on a Petri dish with a 3cm diameter hole, in order to reduce the bottom thickness and optimize visualization of the fluorescence signal. In this way, the only separation between cells and objective is the 90 µm thick coverslip.

**Live-cell imaging.**

Time-lapse phase contrast recordings were acquired with 10X (NA 0.25, air) objective using a Cell Observer Leica DM IRB microscope, images were acquired with a Dalsa DS-21-02M30 CCD Camera by using a 300 seconds interval between acquisitions. A Zeiss Axio Observer Z1, equipped with a Yokogawa CSU-X1A5000 spinning disk confocal scanning unit was also used to acquire fluorescence images of the living cells. The objectives utilized were 10X (NA 0.3, air), 20X (NA 0.4, air), 40X (NA 1.2, water) or 100X (NA 1.4, oil). Images were acquired with a Hamamatsu Orca Flash 4.0 camera. Gamma values were corrected to obtain high contrast images. For the measurements performed with the 10X and 20X objectives images again were recorded with a 300 seconds interval between acquisitions. When using the 40X and the 100X objectives long exposure was not possible because of photobleaching and phototoxicity. Both microscopes were equipped with an on-stage incubation chamber which maintained the temperature at 37°C and $CO_2$ concentration at 5%.

**Data analysis.**

For the analysis of the nuclear size shown in Supplemental Fig. 2 and 4, fluorescence images of our cell line nuclei were automatically thresholded and corrected manually. The following parameters were calculated with Matlab: nuclear area, the length of the minor axis of the ellipse with the same normalized second central moments as the region, and the aspect ratio (i.e. the ratio of the major axis over the minor axis). The frames acquired with Cell Observer and Zeiss Axio Observer were analyzed with ImageJ.

**Optical Stretcher measurements.**

The automated optical stretcher (OS) was used to investigate the mechanical properties of our cell lines in suspended state. In this study we observe cells migrating on a stiff two-dimensional substrate and through rigid quasi-three-dimensional microconstrictions over few hours of times. However, the OS is a useful tool to gain some insight on the cells relative deformability when their actin cytoskeleton is in a depolymerized state (as in the squeezing phase). The main advantage of the OS, compared to other tools to study physical properties of single cells, is that it offers the possibility to manipulate single cells in suspension in absence of any adhesion effect due to the contact with other cells or with the measurement device itself. Prior to measurements, cells cultured at 37°C and 5% $CO_2$ were detached with 0.025% trypsin/EDTA at ~ 80% confluency and transferred to a cryovial after resuspension. The optical fibers of the stretcher were flushed three times with PBS, and the cryovial with cells in suspended state was then connected to the OS fiber system. Cells were then flushed through the fibers in the measurement chamber. Single cells were trapped by two slightly divergent laser beams of a Gaussian profile trap. After 1 s of trapping, the laser power was increased for 2 s (stretching phase), followed by 2 s of further trapping (relaxation phase) (Fig. 4(a)). Deformation was induced on trapped cells by increasing the laser power to 700mW. It was important to measure as many cells as possible, to obtain an accurate broad distribution of the mechanical properties of our specimens. No additional components, such as drugs or dyes, were added to cells. Further details on the experimental apparatus can be found in Morawetz et al. (Morawetz et al., 2017).

Data acquired with the OS were plotted with a Matlab program, written by E. Morawetz. The absence of phase contrast in the dead cells, as well as for the absence of an elastic response in the dead cells after the stretching, allowed the algorithm to sort out dead cells. Focus was adjusted in order to obtain deformation curves independent of any focus effect.

**Simulation.**

Comsol Multiphysics was used to simulate the EGF gradient flow from the culture to the collecting chamber in our microfluidic devices. A two-dimensional approach was used to build a model that closely resembles our system and mimics the features of the microfunnels. The fluid flow field was calculated under stationary conditions. Supplemental figure 1(c) shows the relative concentration of the EGF gradient in the microfluidic device. Since the gradient vanishes in the culture chamber, the focus of the study should be on the region close to the microconstriction row where the gradient flow is present and nearly constant.

**RESULTS**

**Micro-constriction assay.**

The microfluidic device structure introduced in this study is organized in three main layers (Supplemental Fig. 1(a)). The bottom one consists in a 90µm-thick coverslip, which permits the clear visualization of the cells moving inside the micro-funnels with high numerical aperture microscope objectives. The intermediate layer, made of polymerized SU-8 photoresist, defines the cells migration region and creates the micro-funnels. This SU-8 layer also provides the lateral sealing of the chip and ensures a strong adhesion between the top and bottom glass layers. The top layer consists in a 2mm-thick fused silica glass that contains two chambers, one for the seeding of the cells and the other for the injection of the chemoattractant to induce the cells to pass through the constrictions. The funnels are defined by a couple of specular triangular walls (SEM picture reported in Fig. 1(a), in the middle). They are 90 µm long, and are linearly tapered along the cell migration direction from a 40 µm width to a value chosen in the range 3-7 µm (Fig. 1(a) on the right). Many funnels in parallel (18 elements in middle picture in Fig. 1(a)) allow the migration of cells between the two chambers. The height of the SU-8 layer, and thus of the migration region, is 10 µm.

We used three human mammary carcinoma cell lines that well represent the sequential development of cell differentiation: the healthy epithelial MCF-10A, the mesenchymal highly metastatic MDA-MB-231, and the malignant, non-invasive MDA-MB-436 cell line. To create a gradient profile and induce directional migration towards the collecting chamber, we used epithelial growth factor (EGF) as a chemoattractant (1 µl/ml). Cells were loaded in the culture chamber and cultured with regular medium, whereas high-EGF medium was injected in the collecting chamber (Fig. 1(a)). Amoeboid migration has been observed to be more efficient in confined environment or in situations of low adhesiveness (Mak et al., 2016). Accordingly, four-days long phase contrast observations outside and inside the constrictions have revealed that MDA-MB-231 cells seeded on non-coated chips migrated relatively fast with few morphological changes by using coordinated traction forces while responding quickly to changes in external stimuli (Supplemental Video 1). To enhance adhesiveness to the substrate and promote the formation of actin-rich protrusions, the chips were also treated with poly-L-lysine or fibronectin coating solutions. When moving on the two-dimensional surface of the migration region, cells switched between different types of motility modes, depending on surface coating type. On poly-L-lysine or fibronectin-coated surfaces, for instance, the mesenchymal MDA-MB-231 cells maintained a polarized spindle-like morphology, with a leading edge establishing adhesion to the substrate and a contractile rear retracting to allow cellular displacement, whereas on uncoated (glass) surfaces they switched to an amoeboid phenotype. The presence of the microconstriction row interrupted these standard mechanisms of cell motility in 2D. The rigid narrowing constrictions induced a progressive mechanical stress, broadly impacting cell structure and shape and activating significant alterations in the migration pattern. As the funnel width decreased (w <15 µm ≈ cell size, see Supplemental Fig. 2), cells experienced physical confinement by the lateral constricting walls, the bottom and the top layers, favoring the transition to an amoeboid pattern of motility, with roundish morphology and reduced protrusion activity at the leading edge (Supplemental Videos 1-3). Switching motility mode to acquire amoeboid-like characteristics and optimize migration through the microconstrictions was observed in both the MDA-MB-231 and MCF-10A cells, irrespective of surface coating type. In particular, while crossing the micro-funnels, the MDA-MB-231 cells have shown formation of bleb-like projections, more elongated than typical blebs, and radially protruding as branches beyond the funnel opening. In MCF-10A cells, we also observed the formation of lamellipodial extensions spread across the opening trying to establish adhesion to the substrate and pull the cell forward. Cells cultivated on poly-L-lysine or fibronectin coated chips showed no difference in behavior during migration through the microconstrictions and in cell viability when compared to cells cultured on uncoated chips, suggesting that the effects of compression generated by the highly constricting microstructures largely overshadow the effects of adhesion on cell movement in constrictions. The compressive stress enhanced the translocation of highly metastatic MDA-MB-231 cells through the funnels (Fig. 1(c)), and interestingly did not suppress the motility of normal epithelial MCF-10A cells. The epithelial cells responded to the EGF gradient cues by acquiring mesenchymal characteristics (reduced cell-cell contact and persistent motility) and advancing as scattered individual cells, and easily achieved translocation even through the narrowest funnels (Supplemental Video 4). Some cells maintained strong cell-cell junctions and limited motility. The time cells needed to migrate through a microconstriction spanned several hours. The wide time range mainly depended on the specific width of the constriction a cell was stuck into (which here ranges between 3 to 7 µm), as well the variability in cell size and nuclear size. Some cells were able to cross the constrictions in one hour (Supplemental Video 2), whereas other cells could remain stuck up to 12 hours, before successfully reaching the collecting chamber. Interestingly, both mesenchymal and epithelial cells crossed the constrictions in the same time range (approximately 2 hours, see Supplemental Videos 3-4). On the other hand, the malignant, non-metastatic, mesenchymal MDA-MB-436 cells were not particularly responsive to the EGF gradient, exhibited growth arrest and did not migrate through the micro-funnels.

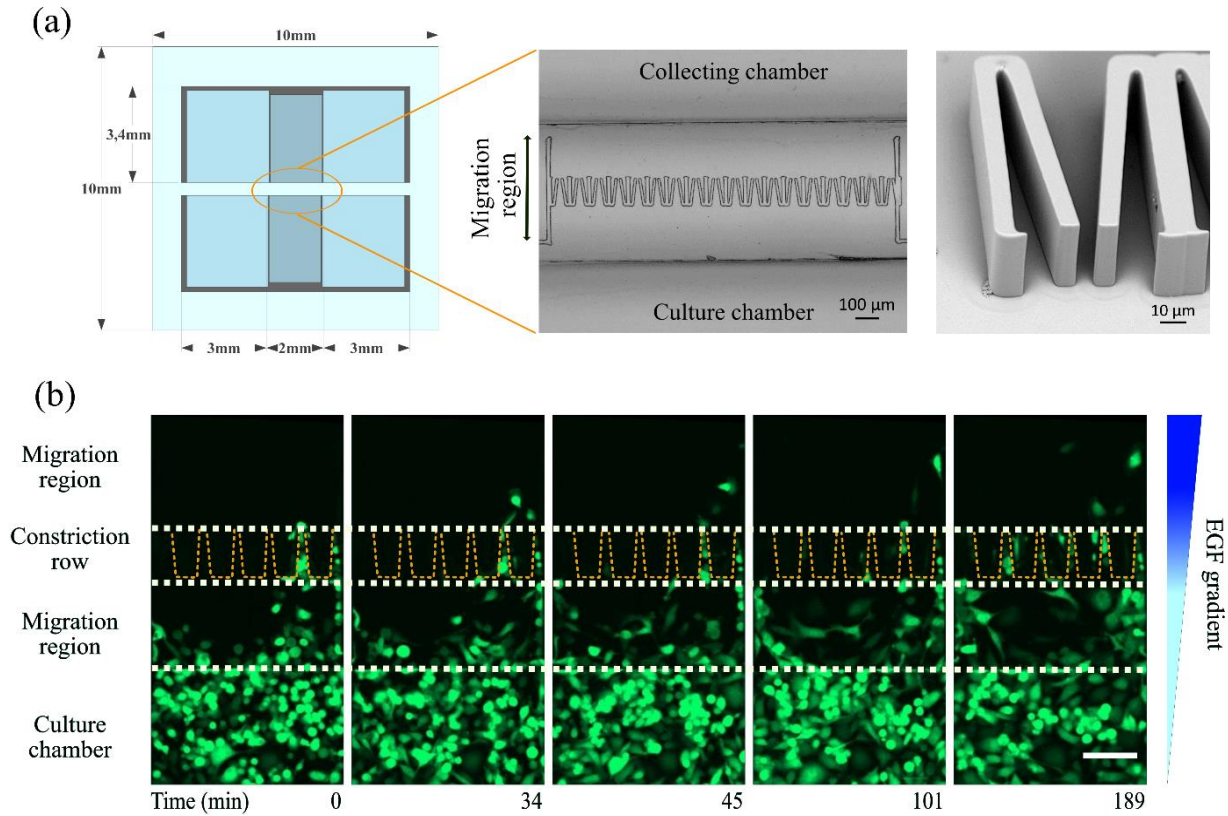

**Figure 1. Overview of micro-constriction chip.**
(a) Schematic image of the micro-funnels device. The two open reservoirs are filled with an equal amount of medium to reduce to minimum any possible hydrostatic pressure imbalance. In the middle image, we use the term migration region to denote a very precise area of the culture (and collecting) chamber, namely the area in proximity to the constriction row and covered by the 2 mm thick silica glass. The right image shows a more detailed view of the micro-funnel row by scanning electron microscopy. (b) Image sequence of MDA-MB-231 cells with stable cytosol expressing GFP migrating through a constriction row. White dashed lines highlight the main sections of the migration region. Yellow dashed lines highlight the shape and position of the microconstrictions. Scale bar is 100 µm.

As previously mentioned in the Methods and Materials section, long-term phase contrast observations were taken over a maximum of four days (upon seeding the cells in the culture chamber). It is important to notice that around the end of the fourth day cells reached high confluence (95%) in the culture chamber and started to crowd the constriction region, not allowing the distinction between cells that were actively migrating through the funnel as individual cells, and those being passively pushed through the micro-constrictions by their neighbours. For this reason, the duration of the experiments was limited to four days, whereas the number of cells observed was limited to a median of 5 MDA-MB-231 cells and a median of 4 MCF-10A cells per single experiments (Supplemental Fig. 3). The experiments were repeated multiple times in order to ensure their reproducibility, as well as the validity of the observations.

**Quasi-three-dimensional confinement induces formation of strongly polarized actin bundles in non-tumorigenic epithelial cells, and favors a transition to a bleb-mediated motility pattern in mesenchymal cells.**

Force generation and changes in cell shape during migration through the constrictions primarily depend on the mechanical behavior of the semi-flexible filamentous actin cytoskeleton, i.e. by the modulation of F-actin crosslinking concentration and filaments length (Stricker et al., 2010). To observe how the dynamical rearrangements of the filamentous actin cytoskeleton relate to the bleb-like cell behavior during the squeezing phase, we labeled the F-actin and acquired higher resolution pictures of cells at different stages of migration.

When moving in the culture chamber and entering the micro-funnels, the MDA-MB-231 cells showed aligned actin stress fibers (Fig. 2, green arrowheads) and a cortical actin shell in proximity of the membrane. Once cells enter the micro-funnels, a stalling phase, which can last up to three hours, was necessary to allow the dynamic depolymerization of the filamentous actin cytoskeleton (Fig. 2, row C). During this stage, cells were stalled at the funnel opening and adapted to the shape imposed by the narrowing structure. Squeezing through the micro-funnel aperture necessitates actin fiber depolymerization, cortical cytoskeleton disruption and deformation of cell membrane. When the cortical cytoskeleton is disrupted at the cell front, the membrane is pushed forward in long bleb-like protrusions (Fig. 3(a)-(b), yellow arrowheads) by cytoplasmic pressure likely generated by myosin II and traction forces of the extracellular adhesion. In strongly adherent and polarized MCF-10A cells, we observed the formation of actin stress fibers at the leading edge during the squeezing phase, suggesting that compression-induced motility increases tension in the actin cytoskeleton (Supplemental Video 5). The newly polarized leading edge elongated to a pronounced lamellipodium to probe the environment and generate sufficient force for translocation of the nucleus through the constriction opening. Overall, the mesenchymal MDA-MB-231 cells and the epithelial MCF-10A cells exhibited fundamentally different migration behaviors when squeezing through the narrowing funnels, with the mesenchymal cells consistently adopting a bleb-based motility mode, and the epithelial cells extending their leading edge across the funnels in lamellipodial projections.

As cells passed through the funnel opening, the nucleus was usually found at the rear of the cell. When it reached a deformation high enough to passively squeeze through the micro-funnel, it bounced beyond the opening, presumably suffering damage at its membrane (Fig. 3(b), white arrowheads). Nevertheless, no morphological indications of apoptotic occurrence, such as apoptotic cell protrusions and cell fragmentation, were detected, and cells preserved their motility after passing through, as confirmed by the long-term phase contrast observations. Compression did not seem to stimulate cell proliferation.

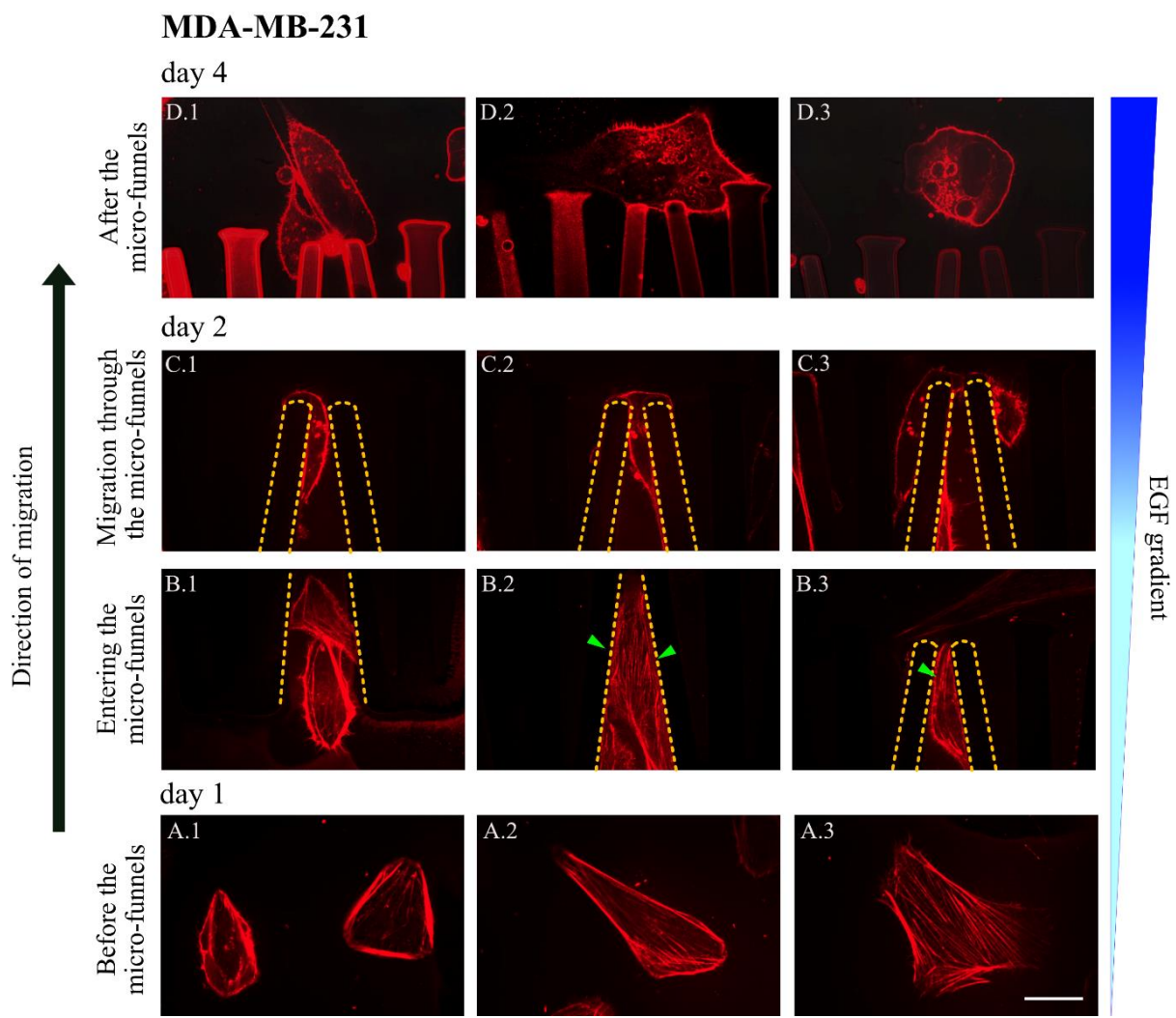

**Figure 2. Migration of MDA-MB-231 cells through micro-constrictions.**
This figure shows the state of different MDA-MB-231 cells during four different stages of migration in the microconstriction region (the cells were observed and tracked during two independent experiments). Starting from below (day 1), row A shows cells migrating towards the miroconstrictions (direction of migration is upward): in the culture chamber, cells display a triangular or ellipsoidal profile and rapid crawling motion with weak substrate adhesion (the glass substrate was not treated). Row B shows cells entering the micro-funnels (day 2): actin stress fibres parallel to the direction of motion are preserved when cells enter the funnels (green arrowheads). Row C shows three cells squeezing through the funnel opening (day 4): disruption of filamentous actin supposedly occurs when cells enter the stalling phase, during which the cytoskeletal structure is reorganized to optimize migration through the funnel openings. During the squeezing phase, cells clearly appear devoid of F-actin, except for a shell in the proximity of the membrane. In row D, three cells have successfully achieved migration through the microconstrictions: cells preserve an irregular shape with an actin shell in proximity of the membrane and no presence of actin bundles. F-actin was labelled in red with SirActin. The constriction profiles are highlighted with yellow dashed lines to improve visualization. Scale bar is 20 µm (for all panels).

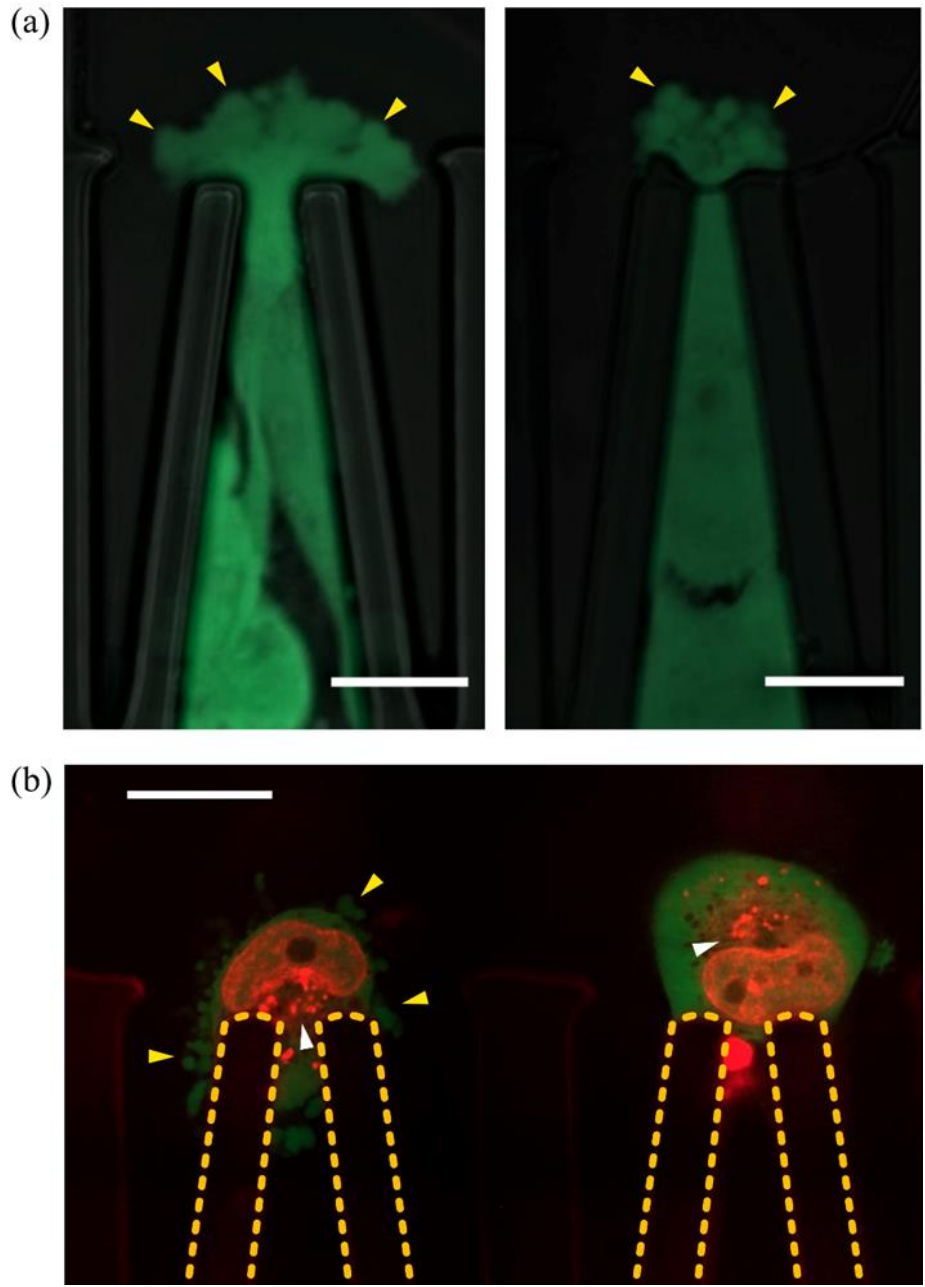

**Fig. 3. Formation of bleb-like structures at the cell front optimize migration through the microconstriction in mesenchymal cells.**
MDA-MB-231 cells with stable cytosol expressing GFP squeezing through a micro-funnel. (a) Extension of bleb-like protrusions is observed when cells try to break free from the surrounding constricting walls. Yellow arrowheads point at the bleb-like protrusions. Sale bar is 20 µm for both pictures. (b) Nuclear squeezing through the highly constrictive funnels may cause nuclear damage in some cells (white arrowheads). The nucleus was stained in red with SirDNA. The constriction profiles are highlighted with yellow dashed lines to improve visualization. Scale bar is 50 µm.

Contrary to the epithelial and invasive cell lines, the malignant MDA-MB-436 cells have in general not shown a response to the EGF gradient, and though a few cells have entered the microconstrictions, they were ultimately not able to squeeze through them. This behavior is at a first sight counterintuitive, especially when considering the results obtained with the chemotactically stimulated epithelial cells, and motivated us to clarify what mechanism may impede fluidization of this type of malignant cell in confinement. The role of cellular stiffness to invasiveness is, to this day, a controversial topic. Previous studies on ovarian cancer cell lines have reinforced the idea that tumor invasiveness is inversely proportional to cellular stiffness, and suggested that determining stiffness of tumor cells can help estimating their metastatic potential (Swaminathan et al., 2011; Xu et al., 2012). On the other hand, a research conducted by Lautscham et al. agrees with the idea that cell stiffness plays only a minor role during migration in 3D micro-environment (Lautscham et al., 2015). By performing experiments with the optical stretcher on all three cell lines, we were able to trap and induce deformation of single cells in suspension and, hence, gain some insight on their relative deformation while maintaining their filamentous actin cytoskeleton in a depolymerized state. The results have confirmed that MCF-10A and MDA-MB-231 cells in suspension are less deformable, and therefore stiffer, than the malignant mesenchymal MDA-MB-436 cells (Fig. 4(a)). Moreover, when comparing the average size of the three cell types (in suspension), the MDA-MB-436 clearly show smaller values, suggesting that cell size is not necessarily a limiting factor in migration through narrow spaces (Fig. 4(b) and Supplemental Fig. 4). Next, we gave a closer look to the structure of the F-actin network of MDA-MB-436 cells during different stages of migration in the migration region. The MDA-MB-436 cells used in this study are characterized in 2D by an extended shape and by the absence of a smooth actin cortex (Fig. 4(c), lower panel). High magnification observations have revealed that inside the micro-funnels these cells lose their elongated morphology, supposedly to adhere more easily to the funnel walls. The images of cells stuck in the micro-funnels show a dense and disorganized F-actin network, but no presence of actin stress fibers that could facilitate the generation of pulling forces. Such cytoskeletal arrangement appears to stiffen the cells and prevent them from behaving as a viscous fluid and squeezing through the microconstrictions (Fig. 4(c), middle and upper panel). Finally, a comparison of the nuclear size of our cell lines exposed in Supplemental

Fig. 2 shows significantly similar values for the MDA-MB-436 and the MDA-MB-231 nuclei, suggesting that successful migration through the narrow channels majorly depends on the cytoskeletal activity, rather than on small nuclear size.

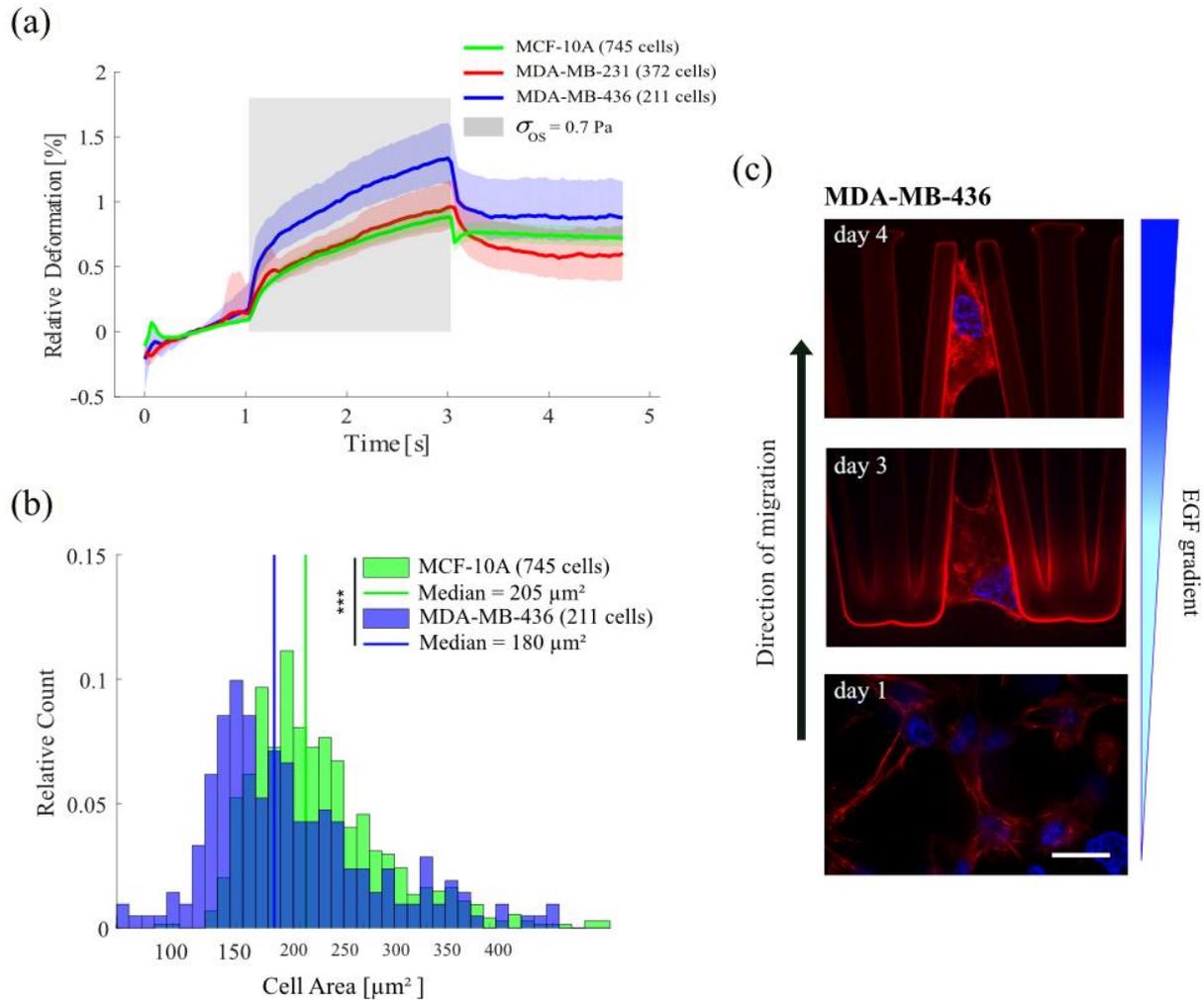

**Fig. 4. Effects of strong confinement on metastatic cells.**
(a) Relative mean deformation of MCF-10A, MDA-MB-231 and MDA-MB-436 cells is plotted. t = 0 – 1 s indicates the trapping phase, where the laser power is used only to trap and stabilize single cells, without stretching them. During the stretching phase t = 1 – 3 s the laser power stepwise increases up to 700 mW and thus induces a stretching force per unit surface of $\sigma_{OS}$ = 0.7 Pa on each individual cell (grey rectangle). MDA-MB-436 are about 50 % more deformable than MDA-MB-231 and MCF-10A cells at the end of stretch (t = 3 s). (b) Histogram of cell area for MCF-10A (green) and MDA-MB436 (blue) cells show similar distribution. Cell area of single suspended cells is calculated during trap phase. Corresponding median (quantile) cell radii are $R_{10A}$ = 8.1 (7.5 8.8) µm and $R_{436}$ = 7.6 (6.8 8.5) µm. MDA-MB-231 express a similar size distribution, see supporting material (Supplemental Fig. 3).

Asterisks represent the level of significance of the two-sided Kolmogorov–Smirnov test with the values $p < 0.05$ (significant), $p < 0.01$ (very significant), $p < 0.001$ (very significant) for one, two and three asterisks, respectively (c) Metastatic MDA-MB-436 cells are shown in two different stages of migration in the migration region. Below panel (day 1): cells adhering to the substrate of the culture chamber have an elongated neuron-like morphology, characterized by a barbed actin profile, no smooth actin cortex, nor aligned actin bundles. Middle and upper panels (day 3 and 4): the inability of metastatic cells to organize the F-actin into stress fibres when entering the narrowing structure prevents them from deforming their shape and squeezing through. Scale bar is 20 µm (for all panels).

**Post-squeezing: Fluid-like behavior impedes recovery of previous shape and mode of migration**.

The compressive stress caused by the highly constricting funnels was not sufficient to trigger cell apoptosis (as evidenced from the fluorescence signal preservation) or arrest cell migration through the microconstrictions, but it was strong enough to impede repolymerization of actin stress fibers within cells, thereby stimulating visible changes in cell shape and cytoskeletal arrangement. In particular, the MDA-MB-231 cells maintained a roundish shape and lacked evidence for mesenchymal traits (as formation of new stress fibers), up to 48 hours after squeezing through the microconstrictions (Fig. 2, row D).

Furthermore, after the squeezing phase, we observed a noteworthy reduction in directional persistence of migration for both MCF-10A and MDA-MB-231 cell lines. Overall, there was a preferential extension during cell migration towards and into the micro-constrictions induced by the chemoattractant gradient, but no preferential extension after migration through the microconstrictions. In the absence of further compression stresses, cells elongated and migrated in random directions (Supplemental Videos 3 and 4).

**EMT is sufficient but not necessary condition for migration of epithelial cells through confining spaces.**

Epithelial cells experiencing neoplastic transformation undergo genetic and epigenetic alterations, that affect tumor suppressor genes and oncogenes. Activation of the transition from epithelial to mesenchymal phenotype, however, is considered as the critical trigger for the acquisition of malignant phenotype by epithelial polarized cells. The transition leads to drastic morphological and molecular changes, while enhancing migratory ability and invasiveness (Chockley and Keshamouni, 2016). The human mammary epithelial MCF-10A cell line is, to this day, considered as a good in vitro model to study transformation and properties of normal epithelial breast cells (Imbalzano et al., 2009; Qu et al., 2015). In our experiments, acquisition of invasiveness in the epithelial MCF-10A manifested itself through persistent single cell migration and invasion through the microconstrictions. More importantly, our data showed that EMT-activated epithelial cells can make their way through the narrowest constrictions with a comparable velocity to that of mesenchymal MDA-MB-231 cells, while preserving their morphological characteristics. This result raised the question whether epithelial cells would still have the capability of fluidizing and successfully achieving migration through rigid confining spaces without employing a chemoattractant.

To determine whether a purely mechanical stimulus could underlie epithelial cell fluidization, we cultured MCF-10A cells in the chip, while both chambers were filled with regular medium. In absence of a chemical gradient, cells proliferated and grew in a monolayer that covered the whole culture chamber within one week. After this time, cells started a collective migration towards the constriction region. Individually migrating cells were not observed, and migration velocity was markedly reduced. The presence of the constriction row manifested itself as a rupture of the confluent state. Surprisingly, the epithelial cells achieved migration through the microconstrictions also in this situation. Migration through the micro-funnels occurred in the same time range as in case of the MCF-10A cells migrating with the EGF gradient, and induced the same alterations to the F-actin system observed previously, though it is not clear whether disruption of the F-actin network was primarily activated by signaling processes in the squeezing cell, or rather by the pushing force generated by the advancing monolayer, or by a combination of the two (Fig. 5). Epithelial cells elongated when pushed into the funnels, but maintained cell-cell adhesion with their neighboring cells during all stages of migration (Fig. 5, green arrowheads). The next 24 hours after squeezing through the funnels, cells preserved some noticeable alterations in their morphology (elongated irregular shape and elongated nuclei, Fig. 6). Cell-cell junctions were still preserved (Fig. 5, green arrowheads), however, migration and settlement in the collecting chamber did not result in the cells forming a new compact monolayer.

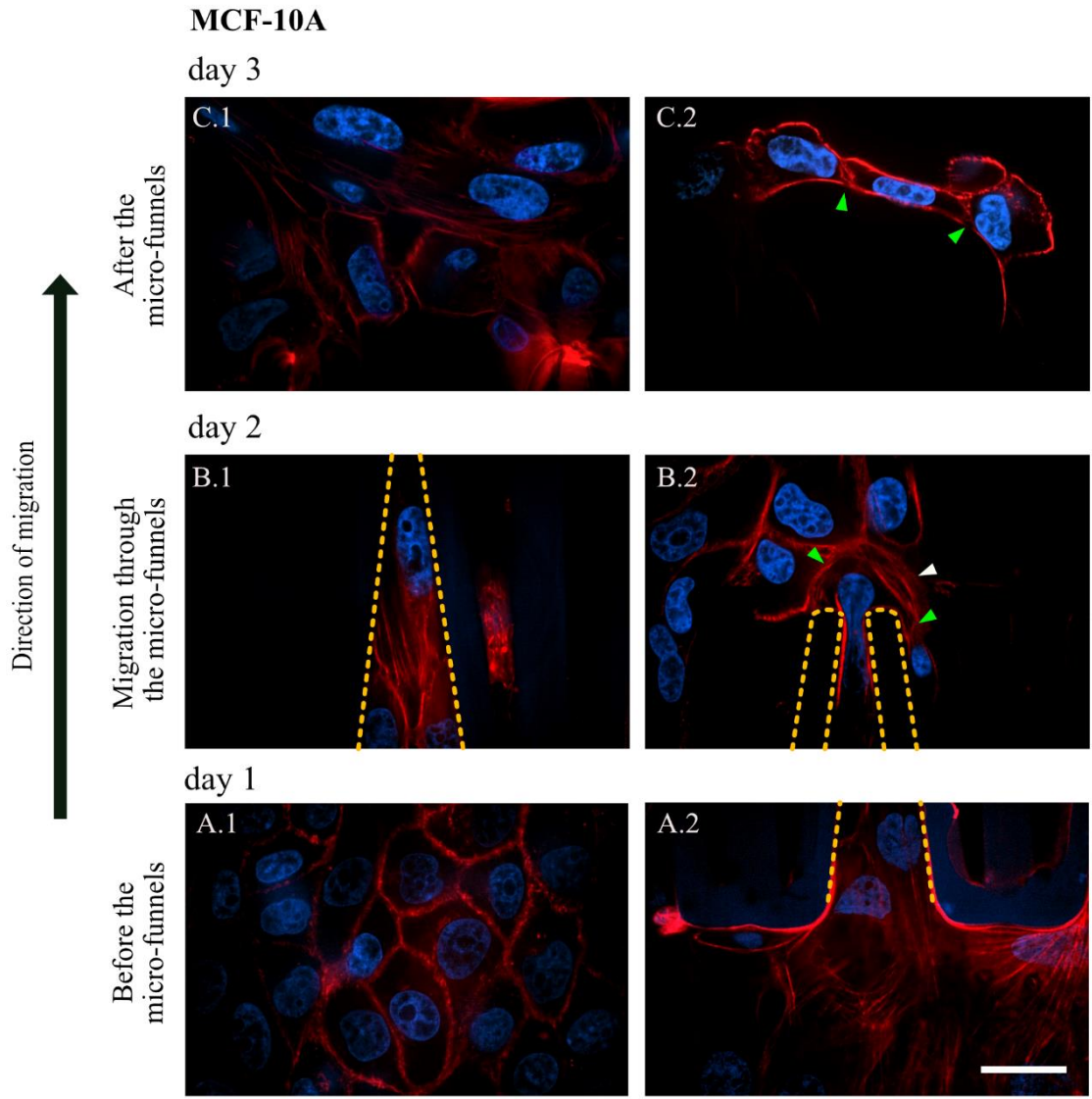

**Fig. 5. Effects of strong confinement on epithelial motile cells in absence of chemoattractive stimulus.**
MCF-10A cells are shown in different phases of migration on an untreated chip. Row A (day 1): without external chemical stimuli, MCF-10A cells proliferate and grow in a compact monolayer. Upon the formation of the compact monolayer, cells start a migration (day 1) towards the constriction row. Cell contact with the funnel wall generates tension in the cytoskeleton and causes the formation of actin stress fibres. Row B (day 2): while being pushed in the microconstrictions, cells form actin bundles aligned to the direction of motion, and subsequently depolymerize them in order to deform their shape and squeeze through the narrow opening. Lamellipodia protrusions (white arrowhead) are used to establish focal adhesion points and induce nuclear squeezing. Row C (day 3): after squeezing through, cells maintain epithelial markers, such as cell-cell adhesion (green arrowheads). The compressive stress experienced by the cells is such that the nuclei maintain an elongated shape after more than 24 hours. For more details about the nuclear shape see Suppl. Fig.4. F-actin is labelled in red with SirActin, the nuclei are labelled in blue with Hoechst. Scale bar is 20 µm (for all panels).

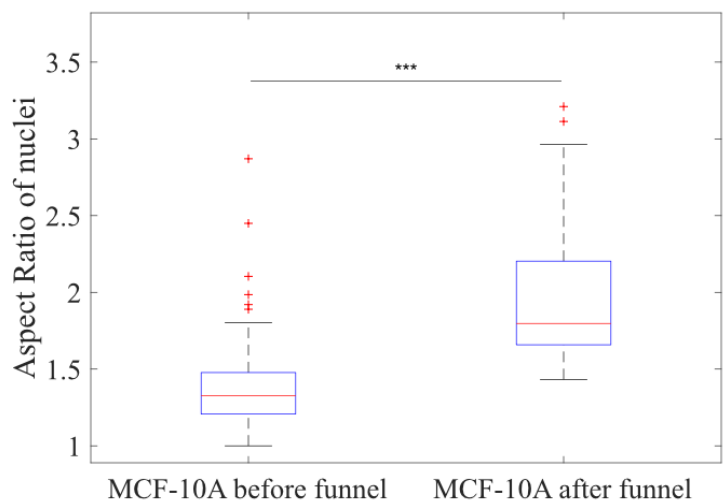

**Fig. 6. Aspect ratio comparison of MCF-10A cells before and after migration through the micro-funnels.** Distribution of the aspect ratio (major axis over minor axis) of the MCF-10A nuclei seeded in the culture chamber and during the post-squeezing phase clearly shows that overall migration through the micro-funnel causes a significant elongation of the nuclei.

**DISCUSSION**

The capability of cells to actively alter their morphology and adapt to the shape imposed by the surrounding micro-environment is a requisite for motile cells to migrate through complex tissues. In living tissues, migrating cells have to apply a considerable stress in order to break free from the surrounding stroma tissue (Friedl and Alexander, 2011). The micro-constriction assay introduced in this paper imposes physical stresses on cells that are chemotactically attracted into rigid micro-funnel structures, thus offering the possibility to observe the response of the filamentous actin cytoskeleton of different breast cancer cell types to a high degree of confinement. Our research wants to elucidate the role of the actin cytoskeleton in cells trapped into compressive rigid micro-channels. We performed cell motility experiments in a novel microfluidic device made of biocompatible materials and non-deformable and smooth surfaces. The device allowed us to observe and compare recurring patterns as well as interesting discrepancies in the migratory behavior of different breast cell types. The narrowing shape and small size of the microconstrictions, in addition to their rigidity, allowed us to observe a limiting case of cell migration in confinement. Crossing the micro-funnels induced disruption of F-actin and formation of membrane protrusions at the cell front. We observed that the epithelial MCF-10A cells and the mesenchymal MDA-MB-231 cells shared the ability to switch between different motility modes in the migration region. The mechanical compression experienced by cells in the micro-funnel induced important alterations in cell shape and cytoskeletal organization. This disruption of filamentous actin did not prevent these cells from advancing through the microconstrictions. Cells adapted their shape to that imposed by the narrowing constrictions, and protruded elongated blebs beyond the constriction opening. Although it may be reasonable to affirm that the mesenchymal MDA-MB-231 cells tend to adopt a bleb-mediated motility pattern when squeezing through the highly constricting funnels, we conclude that this is not the case for the epithelial MCF-10A cells, because of the recurrent and relatively fast formation of dynamic actin-rich lamellipodial-like protrusions at the leading edge during the squeezing phase. Despite maintaining a rounded morphology when they are stuck in the micro-funnels, epithelial cells rely less on blebs and more on lamellipodial mechanism than mesenchymal cells to squeeze the whole cell body through and force nuclear squeezing. Higher resolution observations of the F-actin cytoskeleton of both cell types at different stages of migration from the culture chamber to the collecting chamber revealed the presence of long-term memory effects and the inability of recovering the original cytoskeletal structure.

In 3D migration, cells adopt disparate motility patterns in order to optimize migration through structurally distinct ECMs. Switching motility modes can sometimes be associated with an increase in cell velocity and directional persistence. This is not necessarily the case in our experiments, since cells did not seem to either alter their velocity or maintain directional persistence after crossing the constrictions. Our observations suggest that switching to an amoeboid motility mode in a situation of strong confinement, such as our narrowing microstructures, is not meant to maximize cell speed, but rather facilitates the remodeling of the cytoskeleton while preserving cell integrity throughout the squeezing process. We also showed that confinement does not easily induce either depolymerization, or any phenotypic changes in the malignant mesenchymal MDA-MB-436 cells. We know that the constant interplay of actin binding proteins modulates the assembly and break down of actin filaments into a variety of three-dimensional structures. Transient binding of cross-linking proteins are thought to favor the dynamical rearrangement of the cytoskeletal network and changes in filament density, thus leading to a viscoelastic cell behavior, whereas permanent connections of these proteins are associated with the inability of the cytoskeleton to reorganize itself and behave as an elastic solid (Sun et al., 2010). Although the metastatic cells used in this study were proven to be potentially more deformable than the epithelial MCF-10A cells, they showed minimal response to the chemical and mechanical cues triggered by the soluble chemoattractant and the rigid constriction walls respectively, no disruption of the F-actin cytoskeleton and, consequently, no viscoelastic behavior. The internal mechanisms that prevent cytoskeletal remodeling in this situation, however, are still unclear. This result also indirectly supports our assumption that actin network depolymerization is triggered by signaling cues, rather than representing a mere mechanical breakdown.

Finally, we tested whether EMT is a necessary prerequisite for epithelial cells moving through interstitial spaces, by showing that confinement alone can trigger a transition to a fluid-like behavior in epithelial neoplastic cells, thereby altering their migratory phenotype. Previous research has shown that the normal mammary epithelial MCF-10A cells undergoing mechanical stresses are characterized by high stiffness generated by a dense cytoskeletal network, which would make them less mechanosensitive (Tse et al., 2012). Our observations provide evidence that epithelial cells are just as prone to react to mechanical cues as invasive cells, and have the ability to migrate through narrow spaces without overexpression of the EGF receptor. However, once they reached the collecting chamber, cells preserved an elongated shape and did not restore the original confluent monolayer and tight cell-cell junctions, suggesting that compression-induced fluidization of epithelial cells may have affected the molecular mechanisms underlying cell-cell contacts, while additionally triggering vast changes in cell morphology. Adhesion of same cell type is primarily mediated by two families of molecules, the cadherins and the immunoglobulin (Ig) superfamily. Besides being involved in tissue organization, cadherin adhesion receptors are also implicated in cell differentiation and are thought to be bound to the actin cytoskeleton (Lodish et al., 2000). While binding of cadherin receptors modulates actin filaments, actin assembly and rearrangement of the cortical network boost cadherin clustering. The inability of MCF-10A cells to repair the epithelial monolayer after the rupture of the confluent state may then suggest that the cadherin distribution at the cell surface was altered, resulting in reduced cell-cell adhesion. In other words, the squeezing process may have caused the rupture of some cadherin clusters and the consequent (partial) inactivation of cadherin receptors, while simultaneously stimulating the onset of a different, more mesenchymal mode of migration.

In summary, the microconstriction assay presented in this study allowed us to scan the adaptability of different breast cancer cell types in situation of extreme confinement and detect consistent migration behaviors. Altogether, our results show that compression hugely impacts the cytoskeletal scaffold of motile breast cancer cells, thus inducing quasi-irreversible changes in cell shape and segregation behavior. More importantly, compression experienced by cells inside the micro-funnels did not hinder invasiveness in two fundamentally different cell types, such as the highly aggressive mesenchymal MDA-MB-231 cells and the normal epithelial MCF-10A cells, a result that hardly correlate with the traditional EMT narrative. A thorough understanding of the cytoskeletal dynamics during migration through narrow spaces remains yet to be explored. The contribution of our micro-structures to invasiveness of single cells suggests that this *in vitro* model can be exploited in future for further experimental studies on the role of the cytoskeleton as the predominant active force generator during cancer cell invasion.

**REFERENCES**

Acerbi, I., Cassereau, L., Dean, I., Shi, Q., Au, A., Park, C., Chen, Y., Liphardt, J., Hwang, E., Weaver, V., 2015. Human Breast Cancer Invasion and Aggression Correlates with ECM Stiffening and Immune Cell Infiltration. Integr. Biol. Quant. Biosci. Nano Macro 7, 1120–1134. https://doi.org/10.1039/c5ib00040h

Ahmed, W.W., Betz, T., 2015. Dynamic cross-links tune the solid–fluid behavior of living cells: Fig. 1. Proc. Natl. Acad. Sci. 112, 6527–6528. https://doi.org/10.1073/pnas.1507100112

Amato, L., Gu, Y., Bellini, N., Eaton, S.M., Cerullo, G., Osellame, R., 2012. Integrated three-dimensional filter separates nanoscale from microscale elements in a microfluidic chip. Lab. Chip 12, 1135–1142. https://doi.org/10.1039/c2lc21116e

Bierie, B., Moses, H.L., 2006. TGFβ: the molecular Jekyll and Hyde of cancer: Tumour microenvironment. Nat. Rev. Cancer 6, 506–520. https://doi.org/10.1038/nrc1926

Boucher, Y., Jain, R.K., 1992. Microvascular Pressure Is the Principal Driving Force for Interstitial Hypertension in Solid Tumors: Implications for Vascular Collapse 6.

Brábek, J., Mierke, C.T., Rösel, D., Veselý, P., Fabry, B., 2010. The role of the tissue microenvironment in the regulation of cancer cell motility and invasion. Cell Commun. Signal. 8, 22. https://doi.org/10.1186/1478-811X-8-22

Chockley, P.J., Keshamouni, V.G., 2016. Immunological Consequences of Epithelial–Mesenchymal Transition in Tumor Progression. J. Immunol. 197, 691–698. https://doi.org/10.4049/jimmunol.1600458

Denais, C.M., Gilbert, R.M., Isermann, P., McGregor, A.L., te Lindert, M., Weigelin, B., Davidson, P.M., Friedl, P., Wolf, K., Lammerding, J., 2016. Nuclear envelope rupture and repair during cancer cell migration. Science 352, 353–358. https://doi.org/10.1126/science.aad7297

Friedl, P., Alexander, S., 2011. Cancer Invasion and the Microenvironment: Plasticity and Reciprocity. Cell 147, 992–1009. https://doi.org/10.1016/j.cell.2011.11.016

Imbalzano, K.M., Tatarkova, I., Imbalzano, A.N., Nickerson, J.A., 2009. Increasingly transformed MCF-10A cells have a progressively tumor-like phenotype in three-dimensional basement membrane culture. Cancer Cell Int. 9, 7. https://doi.org/10.1186/1475-2867-9-7

Isermann, P., Lammerding, J., 2017. Consequences of a tight squeeze: Nuclear envelope rupture and repair. Nucleus 8, 268–274. https://doi.org/10.1080/19491034.2017.1292191

Lange, J.R., Fabry, B., 2013. Cell and tissue mechanics in cell migration. Exp. Cell Res., Special Issue: Cell Motility and Mechanics 319, 2418–2423. https://doi.org/10.1016/j.yexcr.2013.04.023

Lautscham, L.A., Kämmerer, C., Lange, J.R., Kolb, T., Mark, C., Schilling, A., Strissel, P.L., Strick, R., Gluth, C., Rowat, A.C., Metzner, C., Fabry, B., 2015. Migration in Confined 3D Environments Is Determined by a Combination of Adhesiveness, Nuclear Volume, Contractility, and Cell Stiffness. Biophys. J. 109, 900–913. https://doi.org/10.1016/j.bpj.2015.07.025

Liu, Y.-J., Le Berre, M., Lautenschlaeger, F., Maiuri, P., Callan-Jones, A., Heuzé, M., Takaki, T., Voituriez, R., Piel, M., 2015. Confinement and Low Adhesion Induce Fast Amoeboid Migration of Slow Mesenchymal Cells. Cell 160, 659–672. https://doi.org/10.1016/j.cell.2015.01.007

Lodish, H., Berk, A., Zipursky, S.L., Matsudaira, P., Baltimore, D., Darnell, J., 2000. Cell-Cell Adhesion and Communication. Mol. Cell Biol. 4th Ed.

Mak, M., Spill, F., Kamm, R.D., Zaman, M.H., 2016. Single-Cell Migration in Complex Microenvironments: Mechanics and Signaling Dynamics. J. Biomech. Eng. 138, 021004. https://doi.org/10.1115/1.4032188

Maruo, S., Fourkas, J.T., 2008. Recent progress in multiphoton microfabrication. Laser Photonics Rev. 2, 100–111. https://doi.org/10.1002/lpor.200710039

Mierke, C.T., 2015. Physical View on the Interactions Between Cancer Cells and the Endothelial Cell Lining During Cancer Cell Transmigration and Invasion. Biophys. Rev. Lett. 10, 1–24. https://doi.org/10.1142/S1793048015300017

Morawetz, E.W., Stange, R., Kießling, T.R., Schnauß, J., Käs, J.A., 2017. Optical stretching in continuous flows. Converg. Sci. Phys. Oncol. 3, 024004. https://doi.org/10.1088/2057-1739/aa6eb1

Park, S.-H., Yang, D.-Y., Lee, K.-S., 2009. Two-photon stereolithography for realizing ultraprecise three-dimensional nano/microdevices. Laser Photonics Rev. 3, 1–11. https://doi.org/10.1002/lpor.200810027

Paul, C.D., Mistriotis, P., Konstantopoulos, K., 2017. Cancer cell motility: lessons from migration in confined spaces. Nat. Rev. Cancer 17, 131–140. https://doi.org/10.1038/nrc.2016.123

Pfeifer, C.R., Xia, Y., Zhu, K., Liu, D., Irianto, J., García, V.M.M., Millán, L.M.S., Niese, B., Harding, S., Deviri, D., Greenberg, R.A., Discher, D.E., 2018. Constricted migration increases

DNA damage and independently represses cell cycle. Mol. Biol. Cell 29, 1948–1962. https://doi.org/10.1091/mbc.E18-02-0079

Qu, Y., Han, B., Yu, Y., Yao, W., Bose, S., Karlan, B.Y., Giuliano, A.E., Cui, X., 2015. Evaluation of MCF10A as a Reliable Model for Normal Human Mammary Epithelial Cells. PLoS ONE 10. https://doi.org/10.1371/journal.pone.0131285

Stricker, J., Falzone, T., Gardel, M.L., 2010. Mechanics of the F-actin cytoskeleton. J. Biomech. 43, 9–14. https://doi.org/10.1016/j.jbiomech.2009.09.003

Sun, S.X., Walcott, S., Wolgemuth, C.W., 2010. Cytoskeletal Cross-linking and Bundling in Motor-Independent Contraction. Curr. Biol. 20, R649–R654. https://doi.org/10.1016/j.cub.2010.07.004

Swaminathan, V., Mythreye, K., O'Brien, E.T., Berchuck, A., Blobe, G.C., Superfine, R., 2011. Mechanical Stiffness Grades Metastatic Potential in Patient Tumor Cells and in Cancer Cell Lines. Cancer Res. 71, 5075–5080. https://doi.org/10.1158/0008-5472.CAN-11-0247

Three-Dimensional Microfabrication Using Two-Photon Polymerization - 1st Edition [WWW Document], n.d. URL https://www.elsevier.com/books/three-dimensional-microfabrication-using-two-photon-polymerization/baldacchini/978-0-323-35321-2 (accessed 9.14.18).

Tse, J.M., Cheng, G., Tyrrell, J.A., Wilcox-Adelman, S.A., Boucher, Y., Jain, R.K., Munn, L.L., 2012. Mechanical compression drives cancer cells toward invasive phenotype. Proc. Natl. Acad. Sci. 109, 911–916. https://doi.org/10.1073/pnas.1118910109

Vishnubhatla, K.C., Bellini, N., Ramponi, R., Cerullo, G., Osellame, R., 2009. Shape control of microchannels fabricated in fused silica by femtosecond laser irradiation and chemical etching. Opt. Express 17, 8685–8695. https://doi.org/10.1364/OE.17.008685

Wu, D., Wu, S.-Z., Xu, J., Niu, L.-G., Midorikawa, K., Sugioka, K., 2014. Hybrid femtosecond laser microfabrication to achieve true 3D glass/polymer composite biochips with multiscale features and high performance: the concept of ship-in-a-bottle biochip. Laser Photonics Rev. 8, 458–467. https://doi.org/10.1002/lpor.201400005

Xu, W., Mezencev, R., Kim, B., Wang, L., McDonald, J., Sulchek, T., 2012. Cell Stiffness Is a Biomarker of the Metastatic Potential of Ovarian Cancer Cells. PLOS ONE 7, e46609. https://doi.org/10.1371/journal.pone.0046609

**Supplemental material.**

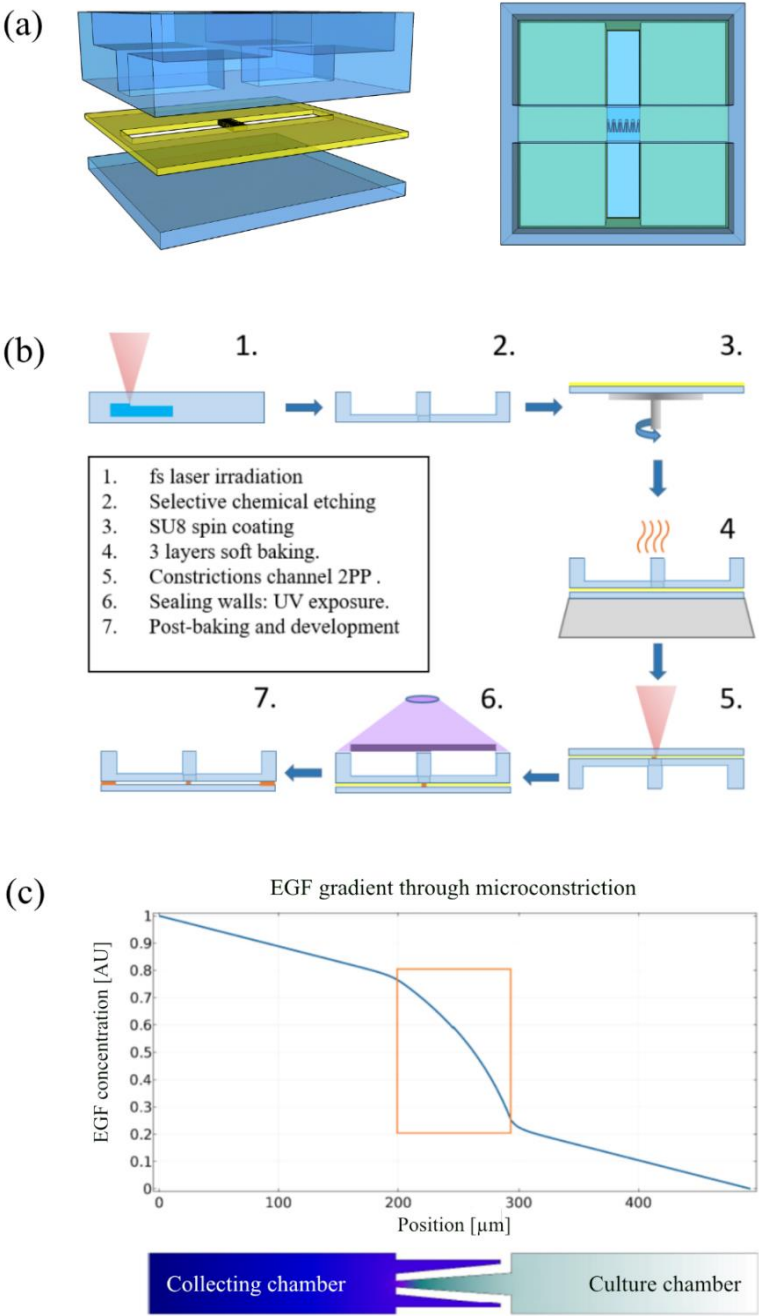

**Supplemental Figure 1. Microfluidic device and EGF gradient profile.** (a) Scheme of the chip for the motility experiment, highlighting its 3 layers, and top view of the finished device. (b) Schematic of the procedure for the micro-chip fabrication. (c) Simulation of diffusion of the EGF-receptor gradient.

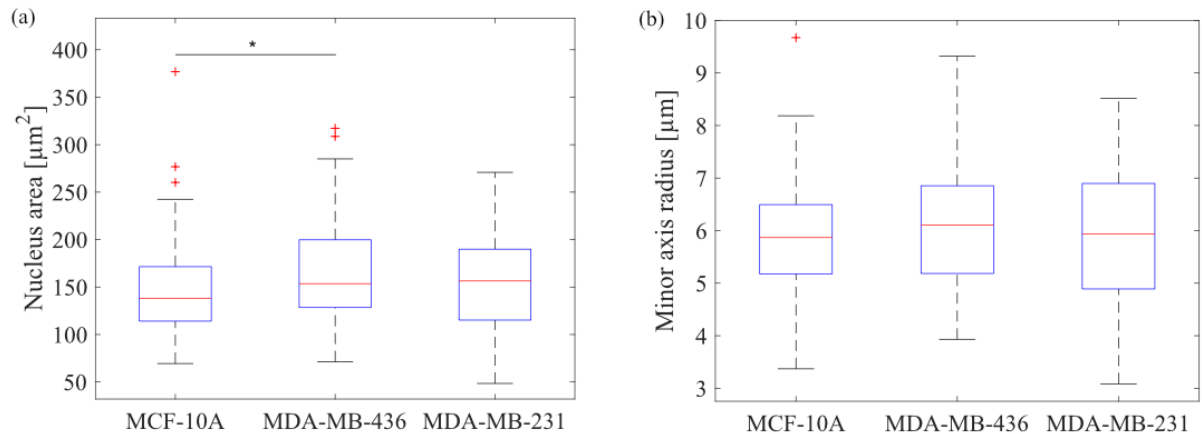

**Supplemental Figure 2. Size comparison of cell nuclei.** Distribution of nuclear area and length of the minor axis radius among our investigated cell lines do not show important differences, as confirmed by the Kolmogorov-Smirnoff test of significance. The asterisk shown in (a) indicates the level of significance of the two-sided Kolmogorov–Smirnov test with $p < 0.05$, $p < 0.01$, $p < 0.001$ for one, two and three asterisks, respectively. As confirmed by the test, the value of the nuclear size of the malignant MDA-MB-436 cells is significantly close to the value of the invasive MDA-MB-231 nuclei, suggesting that the cytoskeleton is the major player during migration through the microfunnels.

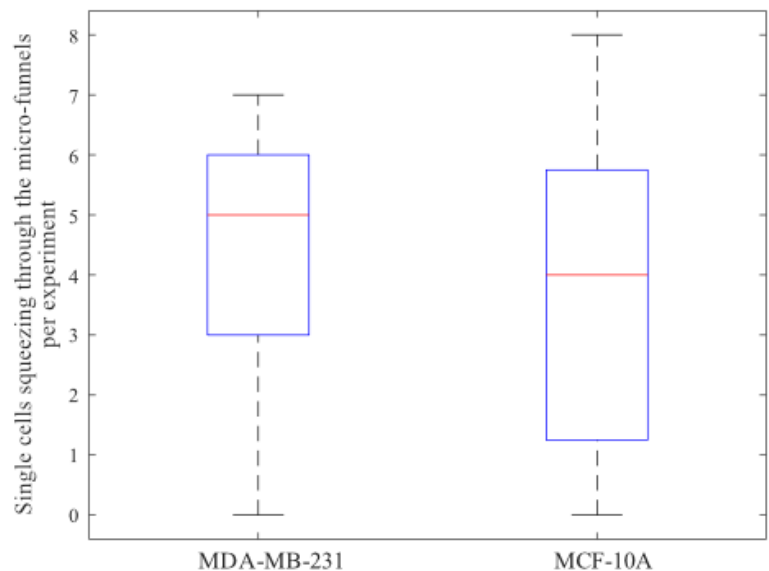

**Supplemental Figure 3. Number of single cells squeezing through the micro-constrictions per experiment.** For the long-term phase contrast observations, a median of 5 cells per single experiment performed with the MDA-MB-231 cells, and a median of 4 for the experiments performed with the MCF-10A cells were found (with a median absolute value of 1 and 2 for the MDA-MB-231 and MCF-10A cells respectively).

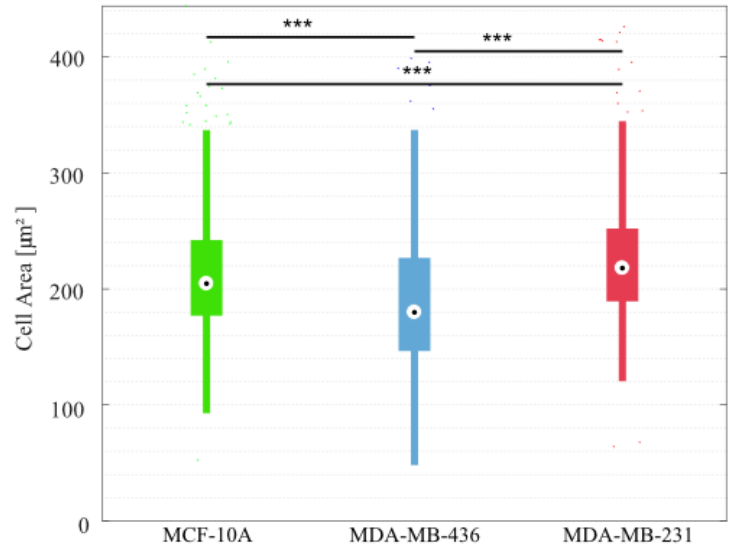

**Supplemental Figure 4. Average cell area.** Distribution of cell size among our investigated cell lines do not show great differences. Median (quantiles) values range from $A_{231}$ = 218 (189 252) µm², $A_{10A}$ = 205 (177 242) µm² to $A_{436}$ = 180 (146 227) µm². Despite the fact that the two-sample Kolmogorov-Smirnov significance test suggests highly different distributions (due to high sample size $n_{10A}$ = 745 cells, $n_{436}$ = 211 cells and $n_{231}$ = 372 cells), we still see a great overlap of the distributions and thus no remarkable difference in cell size of the investigated cell lines.

**Supplemental Video 1. Overview of MDA-MB-231 cells migrating on glass chip.** Phase contrast time lapse imaging of MDA-MB-231 cells migrating on an untreated chip. Cells are observed to adopt an amoeboid motility mode to make their way through the microconstrictions (15 fps, each frame representing 5 min.).

**Supplemental Video 2. Migration of MDA-MB-231 cell through microconstriction – detail #1.** Phase contrast time lapse imaging of MDA-MB-231 cell crossing a micro-funnel (detail from Supplemental video 1). The width of the constriction is 6µm. The cell takes 1h 15min, from the moment the cell senses the funnel opening (15 fps, each frame representing 5 min.).

**Supplemental Video 3. Migration of MDA-MB-231 cell through microconstriction – detail #2.** Phase contrast time lapse imaging of MDA-MB-231 cell crossing a micro-funnel (detail from Supplemental video 1). The width of the funnel opening is 3 µm. Migration through the micro-funnel for this cell lasts two hours, from the moment the cell sense the funnel opening (15 fps, each frame representing 5 min.).

**Supplemental Video 4. Migration of chemotactically stimulated MCF-10A cells through microconstriction** Phase contrast time lapse imaging of MCF-10A cells crossing a microconstriction. Cells migrate individually in the direction of the EGF-gradient and need less than two hours to reach

the collecting chamber. The chip surface was coated with fibronectin, the width of the funnel opening is 3 µm (15 fps, each frame representing 5 min.).

**Supplemental video 5. Formation of actin stress fibers at leading edge of MCF-10A cell squeezing through microconstriction.** Fluorescent time lapse imaging of a MCF-10A cell during the squeezing phase. F-actin was labeled in red with SirActin (15 fps, each frame representing 5 min.).

**Acknowledgements.**

We thank H. Kubitschke for all the advices and helpful discussions, H. M. Scholz-Marggraf for supervising the OS measurements, E. Morawetz for providing the Matlab algorithm to elaborate the OS data, and Jürgen Lippoldt for his help with the analysis of the cell nuclei. This work is supported by the DFG Grant INST 268/296-1 FUGG, and by the ERC Advanced Grant number 741350(HoldCancerBack). The authors declare no competing financial interests.

**Authors contributions.**

J.A.K. and R.O. designed research; C.F. and P.H. performed experiments; C.F. and E.W. performed the OS measurements; C.F analyzed the data; R.M.V., E.L. and J.C. fabricated the chips; C.F. and R.M.V. wrote the paper.